\documentstyle{europhys}
\def\And{{\rm and\ }}

\def\stars{\bigskip\centerline{***}\medskip}

\newif\ifboo \boofalse

\def\Review#1{\boofalse{\it #1},}
\def\Name#1{{\sc #1},}
\def\Vol#1{\ifboo Vol. {\bf #1}\else{\bf #1}\fi}
\def\Year#1{\ifboo #1\else(#1)\fi}
\def\Book#1{\bootrue{\it #1},}
\def\Page#1{\ifboo {\rm p. #1}\else{\rm #1}\fi}
\begin{document}
\euro{}{}{}{}
{\large \bf PREPRINT } 
\Date{\bf accepted for publication in EUROPHYSICS LETTERS}
\shorttitle{R. SAPPEY et. al. : A NEW EXPERIMENTAL PROCEDURE ETC. }
\title{A new experimental procedure for characterizing quantum effects in 
small magnetic particle systems}
\author{R. Sappey\inst{1}\footnote{e-mail address : sappey@spec.saclay.cea.fr}, E. Vincent\inst{1}, M. Ocio\inst{1},  J. Hammann\inst{1}, F. Chaput\inst{2} 
, J.P. Boilot\inst{2}\ And D. Zins\inst{3}}
\institute{
      \inst{1}  Service de Physique de l'Etat Condens\'e, CEA Saclay, 
      91191 Gif-sur-Yvette Cedex, France\\
      \inst{2} Laboratoire P.M.C.,
              CNRS URA D1254,  Ecole Polytechnique, 91128 Palaiseau, France\\
      \inst{3} Laboratoire de Physico-Chimie Inorganique, case courrier 63,
      UPMC, 4 place Jussieu, 75252 Paris, France}
\rec{date1}{in final form date2 }
\pacs{
\Pacs{75}{45$+$j}{Macroscopic quantum phenomena in magnetic systems}
\Pacs{75}{50Tt}{Fine-particle systems}
\Pacs{75}{60Lr}{Magnetic aftereffects}
      }
\maketitle

\input epsf
\epsfverbosetrue

\begin{abstract}
 A new experimental procedure is discussed, which aims at separating thermal
 from quantum behavior independently of the energy barrier distribution in small particle systems. 
 Magnetization relaxation data measured between $60 \ mK$ and $5\  K$ on a sample of 
nanoparticles is presented.  The comparison between experimental data and numerical calculations shows a clear departure from a standard thermal dynamics scenario, a result which was not obvious without using the new procedure presented here.

\end{abstract}

\def\gam{$\gamma-Fe_2O_3$}

The prediction that the magnetic moment of a single domain particle should flip
  by (quasi-macroscopic) quantum tunneling through the anisotropy barrier 
 \cite{Chudnovsky,Enz,Scharf} has motivated many experiments. Some attempts
 at a
 study of  a single  particle  have been  made \cite{Wernsdorfer}, but
most of the experiments have  been carried out on  a macroscopic number of
nanoparticles dispersed  in a non magnetic  matrix, mainly by 
measuring the  magnetic  relaxation  after a  field change (so-called ``viscosity'' measurements) \cite{Tejada}. In such systems  the sizes and  the anisotropy
 constants of the particles are
distributed, and an accurate knowledge of these distributions  is
out of  reach, particularly for the  very small barriers  which are of
interest when measuring the  slow dynamics at low  temperature. In  addition,
there may exist numerous small energy barriers due to  surface defects
 \cite{Wernsdorfer,Berkowitz}, making hazardous a simple correlation between
the energy barriers and the particle sizes. Consequently, the experimental 
evidence of
quantum tunneling of the magnetization (QTM) based on viscosity 
measurements in such systems remains controversial~\cite{Barbara,Gunther}. 
 In this letter, we present relaxation measurements on small isolated
magnetic particles at low temperatures. We describe in the first part the
usual analysis of the thermal  variation of the magnetic viscosity,
emphasizing its limits.  In the second part, we discuss a new
experimental  procedure, which is likely to give  much more reliable information on the thermal or quantum nature of the
observed phenomena.

\section{Viscosity measurements and their limitations}
 The sample   consists   of small  ferrimagnetic  particles   of  \gam
~(maghemite) which  are dispersed in  a  silica matrix,  with a volume
fraction $4.3 \ 10^{-4}$. A  Transmission  Electron Microscopy  study  shows
that  their sizes  can be fitted  to a lognormal  distribution   with peak value
$d_0=6.3\ nm$ and standard deviation  $\sigma=0.25$ \cite{ASI}. It may
thus be considered a good example    of isolated single domain    particle
system.  The  measurements were taken  with a home made combination of
an   r.f.   SQUID     magnetometer and    a     dilution  refrigerator
\cite{thankPP}.  The sample is coupled to the mixing chamber through a
thermal impedance which allows a sample temperature  range of $35\ mK$
to $7  \ K$.  A $  62\ Oe$ magnetic  field is applied  on the sample at an 
initial temperature of $6 \ K$, which proved to produce a well defined 
initial state for subsequent  measurements at lower temperatures; as a
matter of fact, at  $6 \ K$,  all the  processes that one  can  measure at lower
temperatures are rapidly reaching their  equilibrium. This
will appear clearly from the simple calculations below, and has been
checked by other choices than $6 \ K$ \ \cite{ASI}. 
\footnotetext[1]{ By measuring the paramagnetic component of the sample 
signal  \cite{ASI}   , we have  checked   that  the sample temperature
accurately follows the thermometer temperature in the whole accessible
range, with a time constant less than  $60 \ s$.}
\footnotetext[2]{ The time spent before the field cut-off has been checked
to be of no significant influence  on the relaxation rate, in contrast with what has been observed in more concentrated systems, where dipolar interactions
may yield aging effects \cite{Suedois} \ .}
  After field cooling
the sample  down to  the desired  temperature and  waiting for thermal
equilibrium\footnotemark[1], the field is  cut off and the time change
of the  magnetic  moment is recorded by following  the
variation of  the SQUID   signal\footnotemark[2].  This method  avoids
heating problems related  to the sample  movement at low temperatures,
and allows a better sensitivity.  We have studied the variation of the
moment as a function of the time $t$ after the field  cut-off. On one decade of time, between $100 \ s$ and $1000 \ s$,
the plot  is roughly linear in $\log(t)$.  Fig.1 shows the average logarithmic
slope (viscosity $S$)  of   $ M(t)  $  between $100 \ s$ and $1000  \  s$  for
temperatures from $5 \ K$ down to $60 \ mK$.

%%%%%%%%%%%%%%%%%%%%%%%%%% figure 1 %%%%%%%%%%%%%%%%%%%%%%%%%%%%%%%%%%%%%
\begin{figure}[h]
\centerline{\epsfysize=7cm \epsfbox{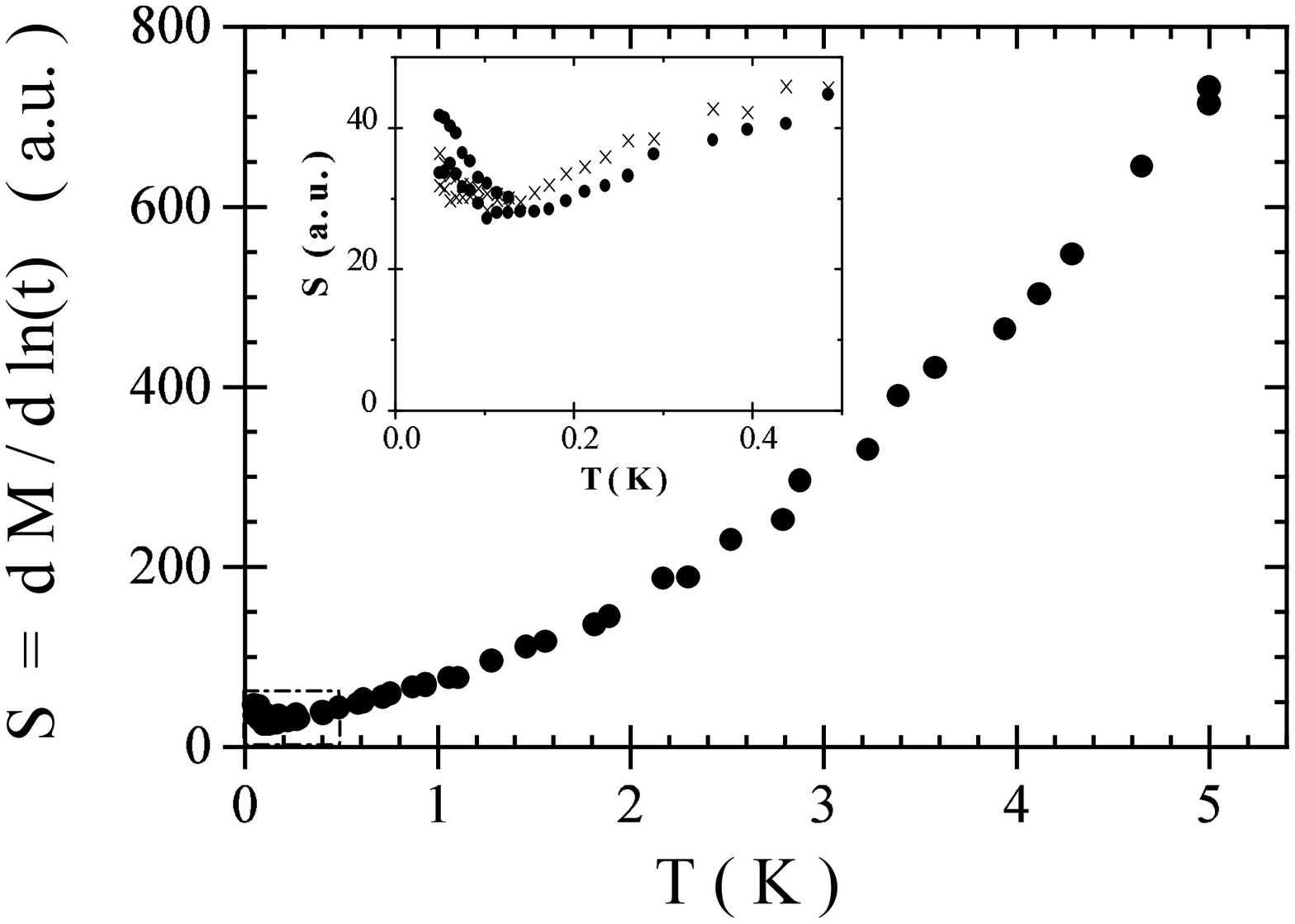}  }

%
%\vbox to 7cm{\vfill\centerline{\fbox{Here is the figure}}\vfill}
%
\caption{Thermal variation of the measured relaxation rate.
A $62\  Oe$ field was cut off at $t = 0 \ s$. Very low T zoom :  $\times$ = rate around $100 \  s$; $\bullet$ =  rate around $1000 \ s$ .}
\label{fig1}
\end{figure}

%%%%%%%%%%%%%%%%%%%%%%%%%%%%%%%%%%%%%%%%%%%%%%%%%%%%%%%%%%%%%%%%%%%%%%%%

For decreasing temperatures, the measured relaxation rate 
first
decreases, then flattens out and surprisingly increases back below 
$150 \ mK $. 
The same result is obtained with a ten times smaller preparation field \cite{ASI} (the viscosity being proportionally reduced). 
 \\
For one isolated uniaxial particle of anisotropy barrier $U$, the relaxation time can be written 
\begin{equation} \label{Tau=...} 
\tau(U,T_0) = \tau_0  \exp \bigg( { U \over k_B T^* (T_0)} \bigg) 
\end{equation}
where $\tau_0$ is a microscopic attempt time of the order of $10^{-8}\ s $ to
$10^{-12}\ s $.
 $T^*(T)$ is an effective temperature, which equals $T$ in the case of
 thermally  activated dynamics. The first-order  predicted effect
 of a crossover towards quantum dynamics is that $T^*(T)$ should become greater
 than $T$  or even temperature independent for temperatures below
a certain crossover temperature $T_{cr}$ \cite{Chudnovsky}, leading to faster 
fluctuations than expected from thermal dynamics.  
 Thinking of the  sample relaxation at temperature $T_0$  as a sum of
independent processes (in our case magnetization reversal of isolated 
particles), one  may   write the total magnetic moment $M(t,T_0)$ as 
\begin{equation} \label{M=sum} M(t,T_0)=\int_0^{+\infty }\ m_{i}(U)\
  P(U)\   \exp \bigg( -{t\over \tau(U,T_0)} \bigg) \ dU 
\end{equation}
 where P(U) is the energy barrier distribution, and $m_{i}(U)$ is the average
 initial magnetic moment of the objects of  anisotropy barrier
U. From eq.\ref{M=sum}  and using the usual step function approximation~\cite{Uehara}, the logarithmic relaxation rate approximates to  
\begin{equation} \label{S=...}
S\equiv \  - \  {\partial M(t,  T_0) \over   \partial   \ln t}\simeq k_B\ 
T^{*}(T_0)\    P(U_c)\ m_{i}(U_c) \   \hbox{, where} \ \    U_c=k_B \ T^{*}(T_0) \ \ln{t\over\tau_0}  \ . 
\end{equation}
 $U_c$ is  the barrier energy of the objects having their main 
contribution to the dynamics after time $t$ at temperature $T_0$. The distribution
$P(U_c) \ m_{i}(U_c)$ has no reason to have a weaker dependence
on $T_0$ than  $T^{*}(T_0)$ itself. The crucial point is that the relevant
part of this distribution  is out of reach in such a system :  at low $T$ and under zero-field, those entities which contribute to the dynamics
correspond to very small energy barriers, whose physical origin remains
uncertain \cite{Wernsdorfer,Berkowitz}. Therefore, extracting any $T^{*}(T_0)$ (which is the quantity of interest to characterize a departure from thermal dynamics)
directly from a measurement of  $S(T)$ is generally not justified, for 
this implies arbitrary hypotheses over $P(U) \ m_{i}(U)$ \ \ 
\cite{Barbara,Gunther,ASI}.  
 In the next section, we present a new experimental procedure 
for reliably distinguishing  between thermal and quantum dynamics, a procedure
which is almost insensitive to the distribution $P(U) \ m_{i}(U)$.

\section{Disentangling thermal from quantum dynamics}
  The point is that the  temperature dependence of the relaxation rate
 is  much  weaker in the  quantum regime  than in the  thermal regime.
 Hence  the following idea  :  the thermal   part   of the   relaxation towards
 equilibrium   should rapidly be  exhausted by   a pre-relaxation at a
 higher temperature, which  in contrast should  be of little influence
 on quantum processes.  Such a procedure is sketched in fig.2a.  After
 field cooling the sample from $6 \ K$ to a temperature $x.T_0$ higher
 than $T_0$, we cut off the field, wait  for $t_0 = 200  \ s$ and then
 cool down the sample to $T_0$. Still considering independent relaxation processes, and neglecting as a first approximation a possible $x$-dependence of the initial moments, we write the relaxation of the total magnetic moment as 
\begin{equation}
 \label{M=...}  M(t,T_0,x) = \int_0^{+\infty  }\ P(U) \ m_i(U) \  \exp
\bigg( -{t_0\over \tau(U,x.T_0)} \bigg)\  \exp \bigg( -{t - t_0  \over
\tau(U,T_0)} \bigg) \ dU \ \ . 
\end{equation}
 As a first step, we now give a qualitative interpretation of our procedure, using the same kind  of  approximation  as in  eq.\ref{S=...}. The
logarithmic slope $S(T_0,x)$ of $M$  around time $1000  \ s$ after the
field cut-off  roughly  corresponds  to the  product of  the  standard
$S(T_0)$ with  a damping factor due  to the pre-relaxation at $x.T_0$.
By  dividing $S(T_0,x)$ by $S(T_0)$,  $P(U_c)\ m_i(U_c)$ is eliminated
(see  eq.\ref{S=...}), and we   obtain a quantity  which  we may  call
``residual memory ratio'' ($RMR$), in the sense that it represents the
memory of the initial state that the system has kept through the whole
procedure. 
\\Beyond this only qualitative argument, we have performed a complete calculation
of eq.\ref{M=...}, and checked
that the variation of this ratio $RMR(x) = S(T_0,x) / S(T_0)$  
indeed contains a very strong information upon $T^{*} (T)$, while being 
remarkably insensitive to the $P(U) \ m_i (U)$ distribution.  We have calculated the time variation of $M$ and its average logarithmic slope
between $\log(t) = 3.0$ and $\log(t) = 3.2$ (this choice has been
 checked  to be of no influence and is related  to  the experimental procedure 
described below). In fig.2b is plotted the calculated  $RMR(x)$ for various 
hypotheses on $T^{*}(T)$ and $P(U) \  m_i(U) $, with  $\tau_0=10^{-10}s$ .
\\In the thermal case (fig.2b curves (i)), $RMR(x)$ is the same for all working 
temperatures $T_0$; in a plateau hypothesis for quantum dynamics (fig.2b curves (ii)), $RMR(x)$ shows a corresponding plateau, followed by a sharp decrease at $x=T_{cr}/T_0$.
%%%%%%%%%%%%%%%%%%%%%%%% figure 2 %%%%%%%%%%%%%%%%%%%%%%%%%%%%%%%%%

\begin{figure}[h]

\centerline{\epsfxsize=6.66cm \epsfbox{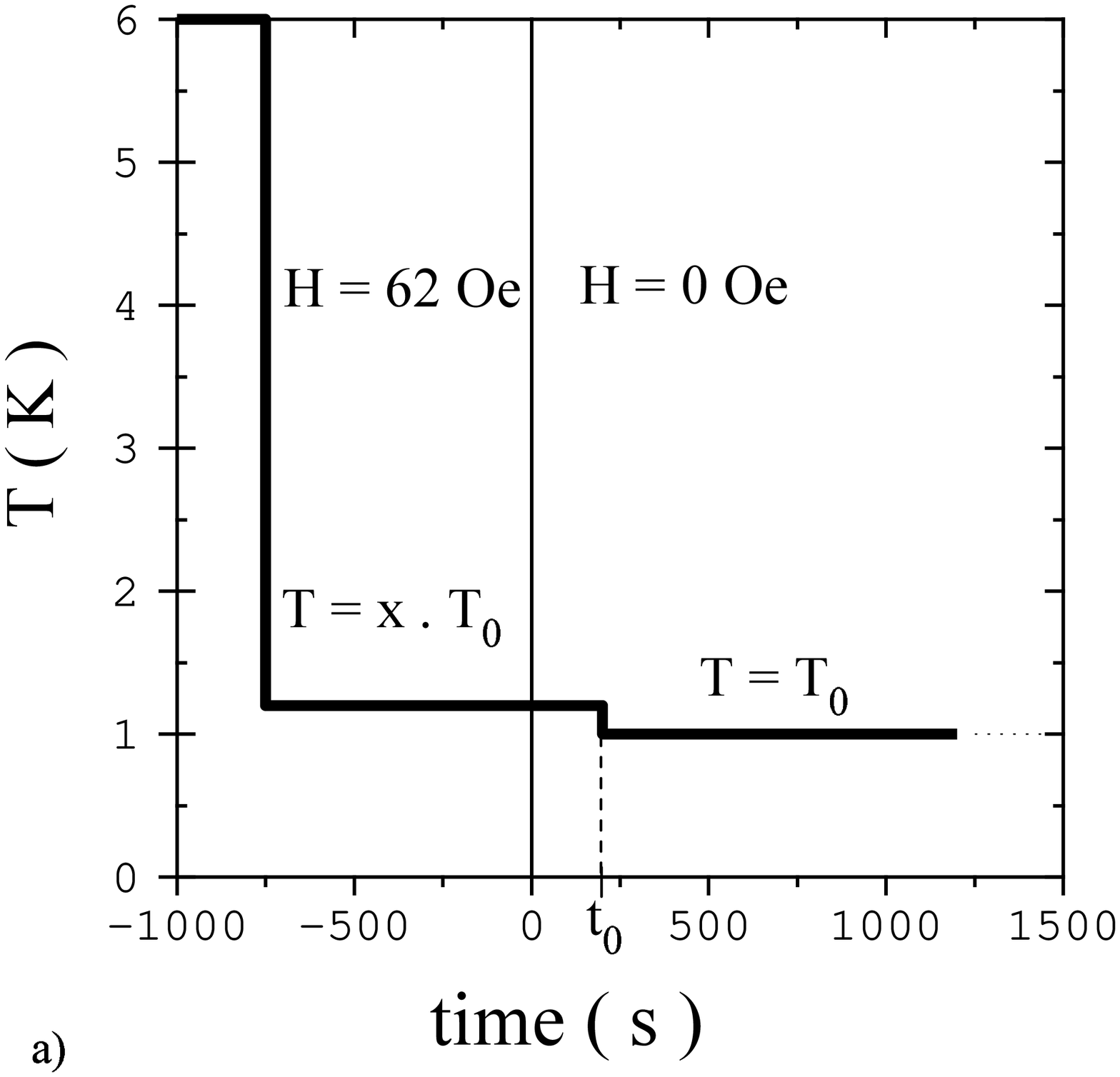} \hskip 0.5cm \epsfxsize=6.66cm \epsfbox{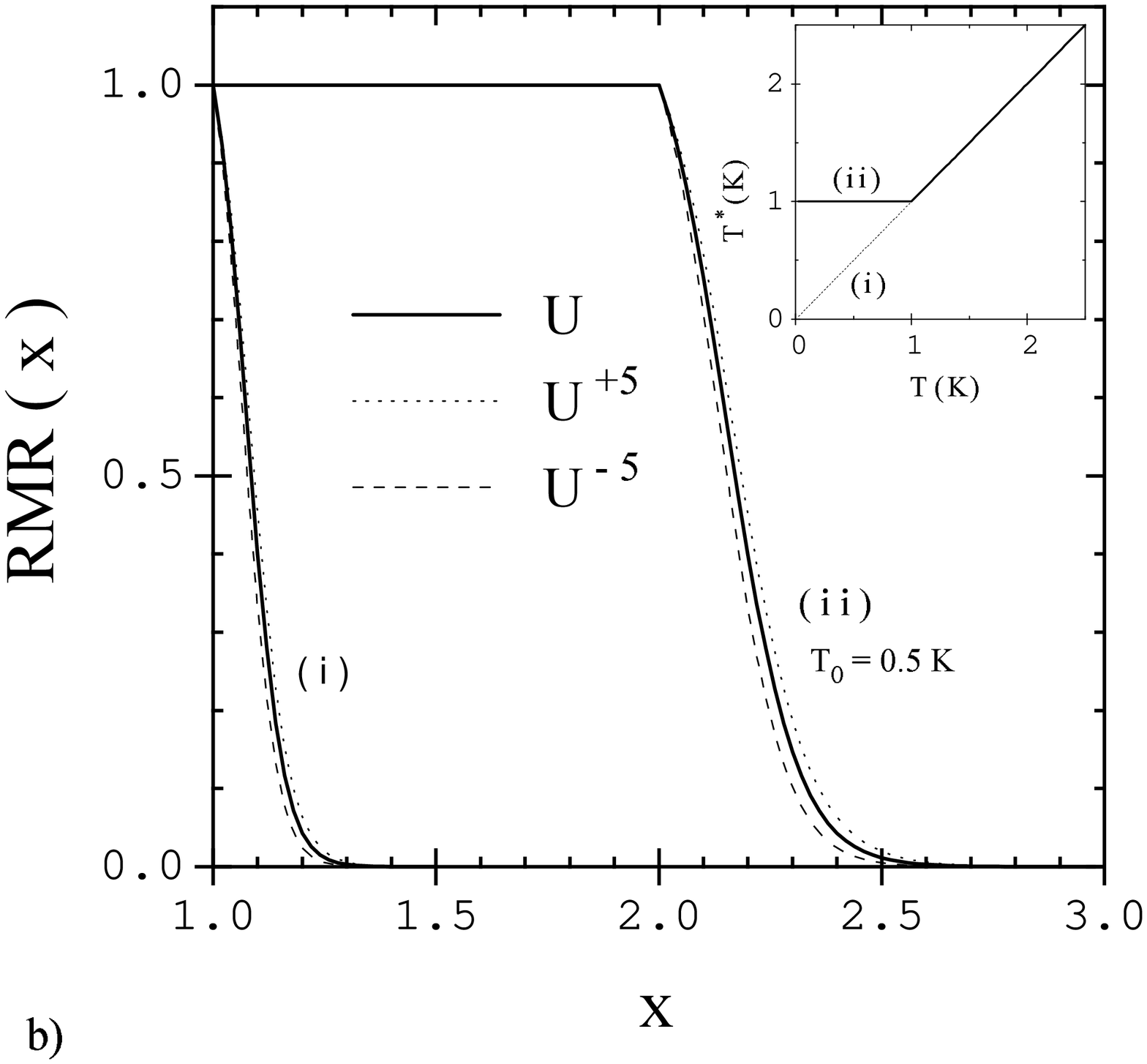} }

%\vbox to 7cm{\vfill\centerline{\fbox{fig2a.ps}}\vfill}

\caption{ a) Typical magnetothermal procedure (here $x = 1.2$). The field is cut off at time $t = 0 \ s$. b)~Calculated $RMR(x)$ for (i) thermally activated
dynamics and (ii) quantum dynamics (in the plateau hypothesis with an
arbitrary choice of $T_{cr} = 1 \ K$ and $T_0 = 0.5 \ K$). The insert shows 
$T^*(T)$ in each case. Three choices of $P(U)\ m_i(U)$ are presented.}
\label{fig2}
\end{figure}

%%%%%%%%%%%%%%%%%%%%%%%%%%%%%%%%%%%%%%%%%%%%%%%%%%%%%%%%%%%%%%%%%%%%%%%%
In both cases, the curves are very weakly dependent on the $\tau_0$ value; for instance, in the thermal case, we have computed that the $x$-value at which $RMR(x)$ has decreased by 90\% ranges from 1.15 for $\tau_0=10^{-8}s$ to 1.20 for $\tau_0=10^{-12}s$. 
\\A crucial result is that, as is clear from fig.2b,  the calculated $RMR(x)$ 
is nearly insensitive to the extremely broad choice of  $P(U) \  m_i(U) $
 presented here (ranging from $U^{-5}$ to $U^{+5}$), while 
it clearly reflects the thermal or quantum nature of the dynamics.
Thus, within our present description, thermal dynamics can be characterized by a sharp decrease of RMR at low $x$. We have assumed temperature independent barriers $U$; in that respect, interactions between the particles must be negligible. This was our motivation for choosing a sample of 
highly diluted particles. Also, we do not consider the case of a singular $P(U) \  m_i(U)$; for instance, in the case of a delta-function peaking at $U \gg U_c$ (see eq.\ref{S=...}), the thermal RMR(x) would exhibit a slower $x$-decrease, but this would imply a very sharp temperature dependence of the viscosity, in disagreement with our present result.
\\ We have applied the procedure sketched in fig.2a to measurements on the same sample as in fig.1; the results are shown in fig.3a.
%%%%%%%%%%%%%%%%%%%%%%%%%% figure 3 %%%%%%%%%%%%%%%%%%%%%%%%%%%%%%%%%%%%%
\begin{figure}[h]

\centerline{\epsfxsize=6.66cm \epsfbox{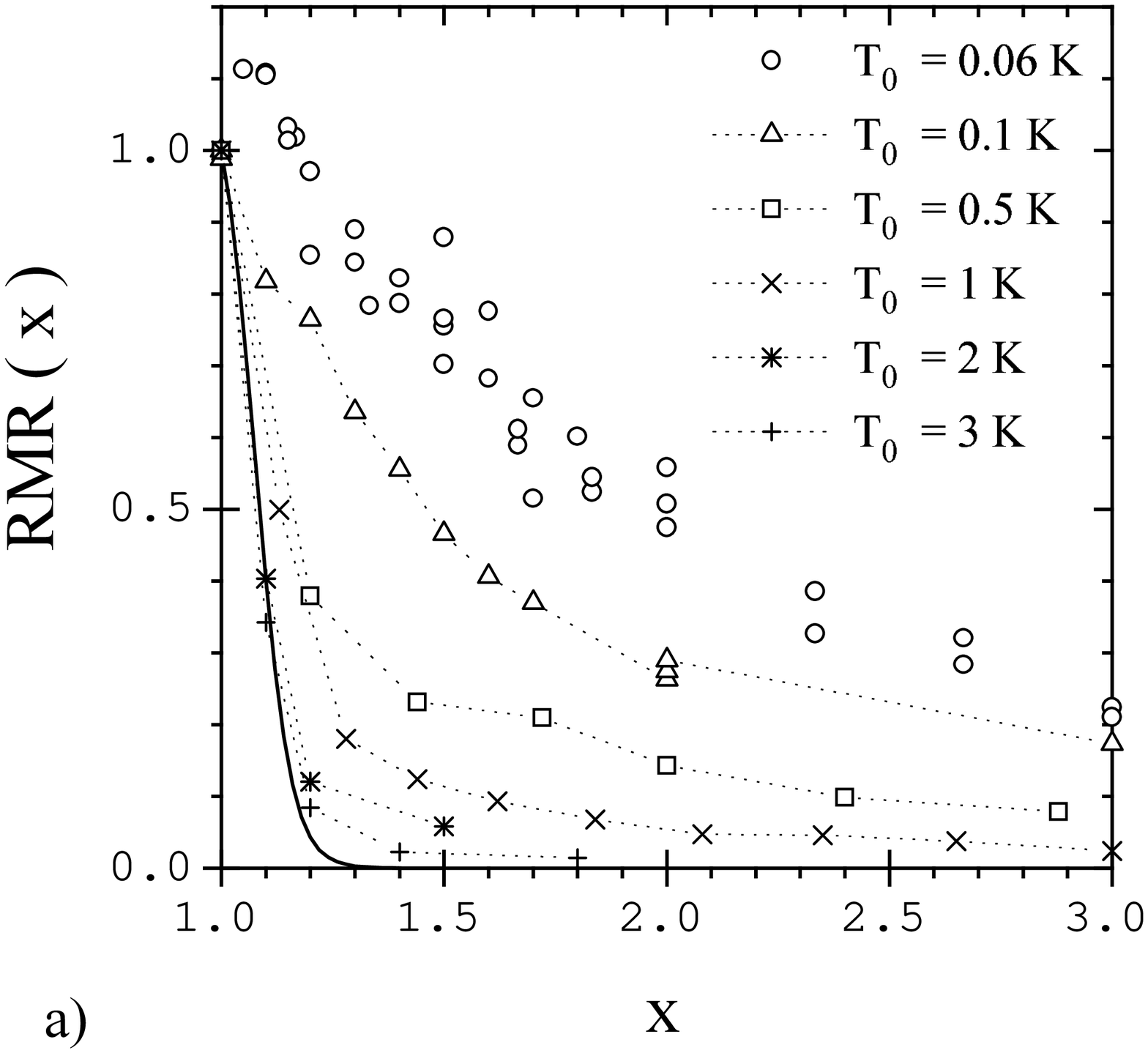} \epsfxsize=6.66cm \epsfbox{fig3bnew.ps} }

%
%\vbox to 7cm{\vfill\centerline{\fbox{Here is the figure}}\vfill}
%
\caption{a) Measured $RMR(x)$ from $3 \ K$ down to $60 \ mK$ (symbols). The 
solid line is the calculation for thermally activated dynamics (see  fig.2b (i)).  b) Calculated $RMR(x)$ for the $T^{*}(T)$ shown in the insert, for different choices of $T_0$.}
\label{fig3}
\end{figure}

%%%%%%%%%%%%%%%%%%%%%%%%%%%%%%%%%%%%%%%%%%%%%%%%%%%%%%%%%%%%%%%%%%%%%%%%
For $T_0 = 3 \ K$ and $T_0 = 2 \ K$, $RMR(x)$ is 
very close to what is expected in the thermal regime. For lower temperatures, 
the observed $RMR(x)$ can no more be explained in terms of thermal dynamics : 
its variation with $x$ becomes slower and slower as temperature decreases, in an intermediate 
fashion between the two extreme cases plotted in fig.2b. The behavior clearly departs from the thermal case of fig.2b(i); however, when compared with the ideal plateau case of fig.2b(ii), the smooth decrease of $RMR(x)$ is
 suggestive of a distribution of crossover temperatures $T_{cr}$ in the system, or of a possible influence of an $x$-dependence of the initial moments. 
\\ Keeping  the same hypotheses as for eq.4, we have computed $RMR(x)$ for a simplistic $T^{*}(T)$ shape as displayed in fig.3b. We have chosen this $T^*(T)$ shape in order to obtain a qualitative agreement with our present experimental data. This $T^{*}(T)$ would still be of about $0.5 \  K$  at our lowest measurement temperature ($0.06 \ K$); 
 this is far beyond any experimental uncertainty on the temperature of our sample. The $T^*(T)$ shape, together with the measured $S(T)$ variation of fig.1,
yields through eq.\ref{S=...} an estimate of the distribution $P(U)\ m_i(U)$. 
It increases towards small
energy barriers, as suggested but not proven in \cite{ASI}. This calls for a better knowledge of the physical origin of low energy  barriers in such samples.
 \\ \\ As a conclusion, this new procedure brings significant  information
on the nature of the dynamics, almost independently of the magnetic moment 
distribution  $P(U)\ m_i(U)$. The method has allowed us to evidence a 
significant  departure from the expected thermal dynamics scenario in an assembly of small magnetic particles.
 In our opinion, this method should help evidencing and studying QTM in 
many-particle samples, and could be interestingly extended to the characterization of non-thermal behavior in other systems,
such as depinning of vortices in superconductors or of Bloch walls in ferromagnets.

\stars


\vskip-12pt

\begin{thebibliography}{99}

\bibitem{Chudnovsky}
\Name{Chudnovsky E. \And Gunther L.} \Review{Phys. Rev. Lett.} \Vol{60} \Year{1988} \Page{661}.

\bibitem{Enz}
\Name{Enz M. \And Schilling R.} \Review{J. Phys. C} \Vol{19} 
\Year{1986} \Page{L711}

\bibitem{Scharf}
\Name{Scharf G., Wreszinski W. F., \And van Hemmen J. L.}
\Review{J. Phys. A} \Year{1987} \Page{4309}

%\bibitem{Wernsdorfer}
%\Name{Wernsdorfer W. , Hasselbach K. , Mailly D. ,  Barbara B. , Benoit A. , Thomas A. \And Suran G. } \Review{J. Magn. Magn. Mater.} \Vol{145} \Year{1995} 
%\Page{33}.

\bibitem{Wernsdorfer}
\Name{Wernsdorfer W. , Hasselbach K. , Benoit A. , Cernicchiaro G. , Mailly D.,
  Barbara~B. , \And L.~Thomas} \Review{J. Magn. Magn. Mater.} \Vol{151} 
\Year{1995} \Page{38}.



\bibitem{Tejada}
\Name{Tejada J. , Zhang X.~X. , \And Balcells L.} \Review{J. Appl. Phys.} 
\Vol{73} \Year{1993} \Page{6709}.


\bibitem{Berkowitz}
\Name{Kodama R. H. , Berkowitz A. E. ,  McNiff Jr. E. J. , \And Foner S.} 
\Review{Phys. Rev. Lett.} \Vol{77} \Year{1996} \Page{394}.

\bibitem{Barbara}
\Name{Barbara B. , Sampaio L. C. , Marchand A. , Kubo O. , \And Takeuchi H. }
 \Review{J. Magn. Magn. Mater.} \Vol{136} \Year{1994} \Page{183}.

\bibitem{Gunther}
\Name{Gunther L.} in \Book{Quantum Tunneling of Magnetization 
- QTM'94, Chichilianne 1994}, edited by \Name{L. Gunther \And B. Barbara}, ( Kluwer Academic Publishers ) \Year{1995} \Page{413-434}.

\bibitem{ASI}
\Name{Sappey R., Vincent E. , Hammann J. , Chaput F. , Boilot J.~P. , 
and Zins D.} to be published in \Book{Magnetic Hysteresis in Novel Magnetic Materials, Greece 1996}, edited by \Name{H. J. Hadjipanayis}, ( Kluwer Academic Publishers ).
\newblock Preprint cond-mat/9609025 available on http://xxx.lanl.gov.

\bibitem{Suedois}
\Name{Jonsson T., Mattsson J., Djurberg C., Khan F. A., Nordblad P. \And
Svedlindh P.} \Review{Phys. Rev. Lett.} \Vol{75} \Year{1995} \Page{4138}

\bibitem{thankPP}
\newblock The special-purpose dilution refrigerator has been designed and built
  in CEA-Saclay by P. Pari and Ph. Forget, with the help of L. Le Pape.

\bibitem{Uehara}
\Name{Uehara M. \And Barbara B.} \Review{J. Physique} \Vol{47} \Year{1986}
\Page{235}.




\end{thebibliography}
\end{document}